\documentclass{appolb}
\usepackage{epsfig}
\begin{document}
\title{$N^*$ and Meson Resonances in $J/\psi$ decays%
\thanks{Presented at the Meson2002, 7th International Workshop on 
Production,
Properties and Interaction of Mesons, Cracow, Poland, May 23-28, 2002 }%
}
\author{Bing-song Zou, 
\address{
Institute of High Energy Physics, Chinese Academy of Sciences,\\
     P.O.Box 918 (4), Beijing 100039, P.R.China\\\medskip
(representing BES Collaboration)\\
}}
\maketitle
\begin{abstract}
Over sixty million $J/\psi$ events have been collected by the BES 
Collaboration at
the Beijing Electron-Positron Collider (BEPC). $J/\psi$ decays provide
an excellent place for studying excited nucleons and hyperons -- $N^*$,
$\Lambda^*$, $\Sigma^*$ and $\Xi^*$ resonances, as well as meson resonances,
including possible glueballs and hybrids. Physics objectives, recent results
and future prospects of light hadron spectroscopy at BEPC are presented.
\end{abstract}
\PACS{13.25Gv, 14.40Cs, 14.20Gk, 13.65+i}
  
\section{Introduction}

The Institute of High Energy Physics at Beijing runs an electron-positron
collider (BEPC) with a general purpose solenoidal detector,
the BEijing Spectrometer (BES)\cite{BES},
which is designed to study exclusive final states in $e^+e^-$
annihilations at the center of mass energy from 2000 to 5600 MeV.
In this energy range, the largest cross sections are at the $J/\psi(3097)$ 
and $\psi'(3686)$ resonant peaks.
Up to now, the BES has collected about 65 million $J/\psi$ events and 
18 million $\psi'$ events. From $J/\psi$ and $\psi'$ decays, both meson
spectroscopy and baryon spectroscopy can be studied.

Three main processes which play very important role for the light hadron
spectroscopy are $\psi$ hadronic decay into baryons and anti-baryons,
$\psi$ radiative decay, and $\psi$ hadronic decay into mesons. 
In the following three sections, I will outline the
physics objectives and summarize recent results for each of them.
Future prospects are given in the final section.

\section{$N^*$ and hyperons from $J/\psi$ decays}

Baryons are the basic building blocks of our world. If we cut any piece
of object smaller and smaller, we will finally reach the nucleons,
{\sl i.e.}, the lightest baryons, and we cannot cut them smaller any
further. So without mention any theory, we know that the study of baryon
structure is at the forefront of exploring microscopic structure of
matter.  From theoretical point of view,  since baryons represent the
simplest system in which the three colors of QCD neutralize into
colorless objects and the essential non-Abelian character of QCD is
manifest, understanding the baryon structure is absolutely necessary
before we claim that we really understand QCD.

Spectroscopy has long proved to be a powerful tool for exploring internal
structures and basic interactions of microscopic world. Ninety years ago
detailed studies of atomic spectroscopy resulted in the great
discovery of Niels Bohr's atomic quantum theory\cite{Bohr}. Forty to sixty
years later, still detailed studies of nuclear spectroscopy resulted in
Nobel Prize winning discoveries of nuclear shell model\cite{SM} and
collective motion model\cite{CM} by Aage Bohr {\sl et al}.
Comparing with the atomic and nuclear spectroscopy at those times,  
our present baryon spectroscopy is still in  its infancy\cite{PDG}.
Many fundamental issues in baryon spectroscopy are still not well
understood\cite{Capstick1}. The possibility of new, as yet unappreciated,
symmetries could be addressed with accumulation of more data. The new   
symmetries may not have obvious relation with QCD, just like nuclear
shell model and collective motion model.

\begin{figure}[htbp]
\vspace{-0.8cm}
\hspace{0.5cm}\includegraphics[width=14cm,height=5.5cm]{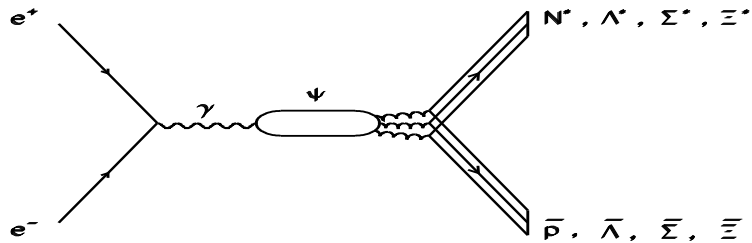}
\vspace{-1.1cm}
\caption{$\bar pN^*$, $\bar\Lambda\Lambda^*$,
$\bar\Sigma\Sigma^*$ and $\bar\Xi\Xi^*$ production
from $e^+e^-$ collision through $\psi$ meson.}
\label{fig:1}
\end{figure}

Joining the new effort on studying the excited nucleons, $N^*$ baryons,
at new facilities such as CEBAF at JLAB, ELSA at Bonn, GRAAL at Grenoble
and SPRING8 at JASRI, we also started a baryon resonance program at 
BES\cite{Zou1}, at Beijing Electron-Positron Collider (BEPC).
The $J/\psi$ and $\psi'$ experiments at BES provide an
excellent place for studying excited nucleons and hyperons -- $N^*$,
$\Lambda^*$, $\Sigma^*$ and $\Xi^*$ resonances\cite{Zou2}.
The corresponding Feynman graph for the production of these excited 
nucleons and hyperons is shown in Fig.~\ref{fig:1} where $\psi$
represents either $J/\psi$ or $\psi'$.

Comparing with other facilities, our baryon program has advantages in at
least three obvious aspects:

(1) We have pure isospin 1/2 $\pi N$ and $\pi\pi N$ systems from
$J/\psi\to\bar NN\pi$ and $\bar NN\pi\pi$ processes due to isospin   
conservation, while $\pi N$ and $\pi\pi N$ systems from $\pi N$ and    
$\gamma N$ experiments are mixture of isospin 1/2 and 3/2, and suffer  
difficulty on the isospin decomposition;

(2) $\psi$ mesons decay to baryon-antibaryon pairs through three or
more gluons. It is a favorable place for producing hybrid (qqqg) baryons,
and for looking for some ``missing" $N^*$ resonances which have weak
coupling to both $\pi N$ and $\gamma N$, but stronger coupling to $g^3N$;

(3) Not only $N^*$, $\Lambda^*$, $\Sigma^*$ baryons,
but also $\Xi^*$ baryons with two strange quarks can be studied.       
Many QCD-inspired models\cite{Isgur,Glozman} are expected to be more   
reliable for baryons with two strange quarks due to their heavier quark
mass. More than thirty $\Xi^*$ resonances are predicted where only
two such states are well established by experiments. The theory is totally
not challenged due to lack of data.

BES started data-taking in 1989 and was
upgraded in 1998. The upgraded BES is named BESII while the previous one
is called BESI. BESI collected 7.8 million $J/\psi$ events and 3.7 million
$\psi'$ events. BESII has collected 58 million $J/\psi$ events.

\begin{figure}[htbp]
\centerline{ \psfig{file=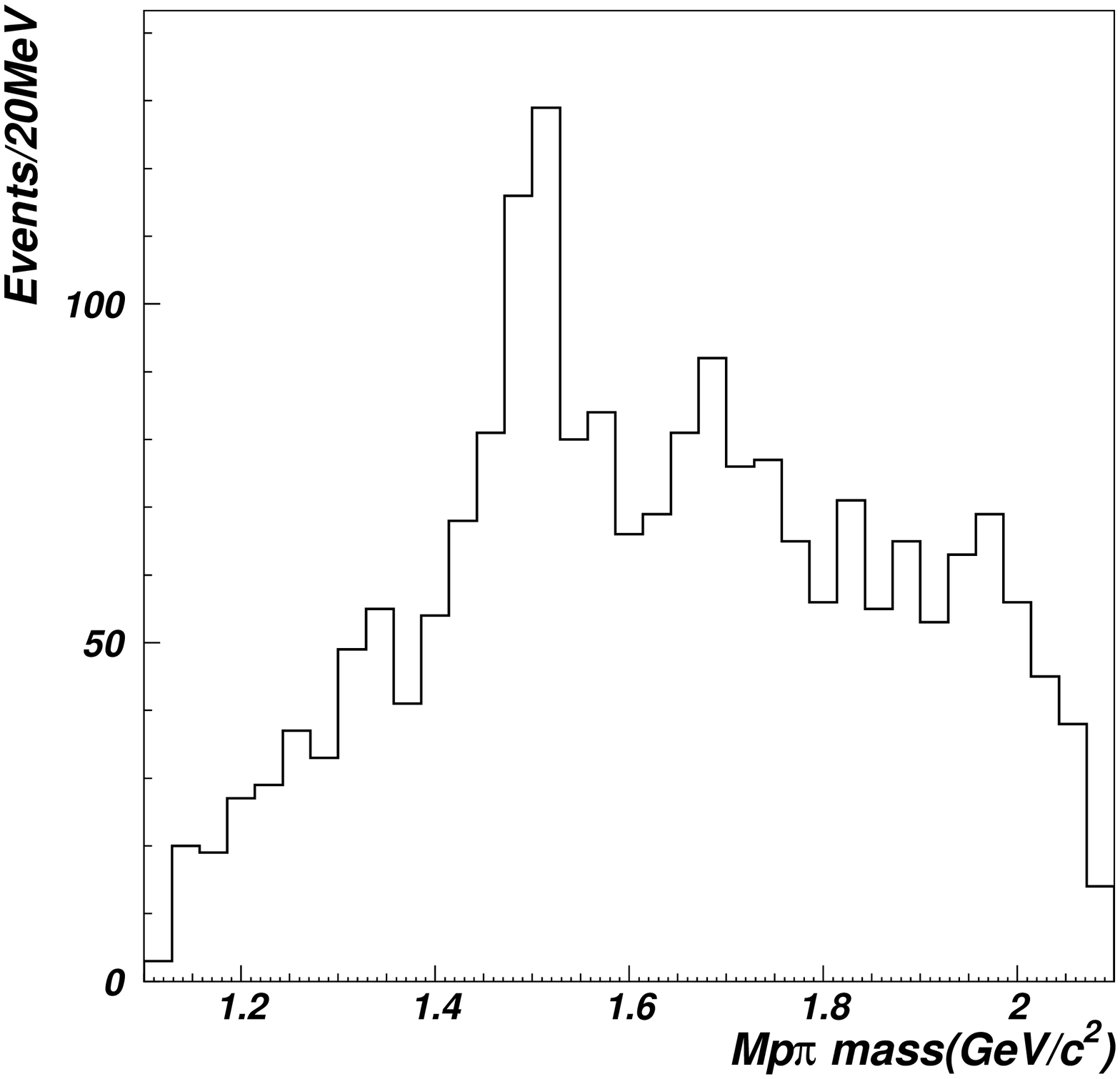,height=6.5cm,angle=0,silent=}
\hspace{-0.2cm} \psfig{file=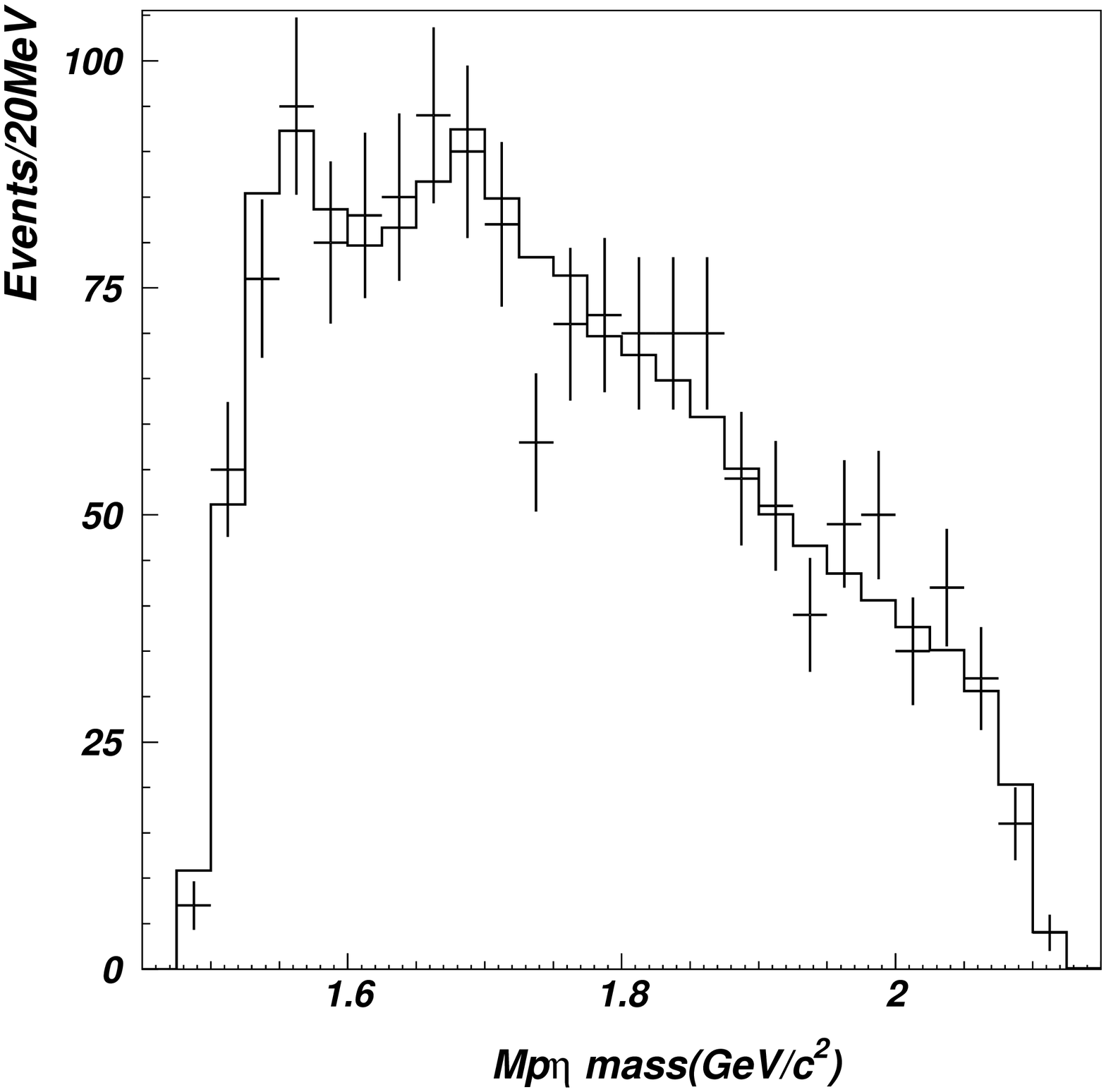,height=6.5cm,angle=0,silent=}
}
\caption{\label{fig2} left: $p\pi^0$ invariant mass spectrum for
$J/\psi\to\bar pp\pi^0$; right: $p\eta$ invariant mass spectrum for
$J/\psi\to\bar pp\eta$. BESI data}
\end{figure}

Based on 7.8 million $J/\psi$ events collected at BESI before 1996,
the events for $J/\psi\to\bar pp\pi^0$ and $\bar pp\eta$ have been
selected and reconstructed with $\pi^0$ and $\eta$ detected in their 
$\gamma\gamma$ decay mode\cite{Zou1}.
The corresponding $p\pi^0$ and $p\eta$ invariant mass spectra are
shown in Fig.~\ref{fig2} with clear peaks around 1500 and
1670 MeV for $p\pi^0$ and clear enhancement around the $p\eta$ threshold,
peaks at 1540 and 1650 MeV for $p\eta$.
Partial wave analysis has been performed for the $J/\psi\to\bar pp\eta$
channel\cite{Zou1}
using the effective Lagrangian approach\cite{Nimai,Olsson} with Rarita-Schwinger
formalism\cite{Rarita,Fronsdal,Chung,Liang} and the extended automatic
Feynman Diagram Calculation (FDC) package\cite{Wang}.
There is a definite requirement for a
$J^{P}=\frac{1}{2}^-$ component at
$M = 1530\pm 10$ MeV with $\Gamma =95\pm 25$ MeV near the $\eta N$
threshold. In addition, there is an obvious resonance around 1650 MeV
with $J^P=\frac{1}{2}^-$ preferred, $M = 1647\pm 20$ MeV and
$\Gamma = 145^{+80}_{-45}$ MeV.
These two $N^*$ resonances are believed to be the two
well established states, $S_{11}(1535)$ and $S_{11}(1650)$, respectively.
In the higher $p\eta$($\bar{p}\eta$) mass
region, there is a evidence for a structure around 1800 MeV;
with BESI statistics we cannot determine its quantum numbers.

\begin{figure}[htbp]
\centerline{ \psfig{file=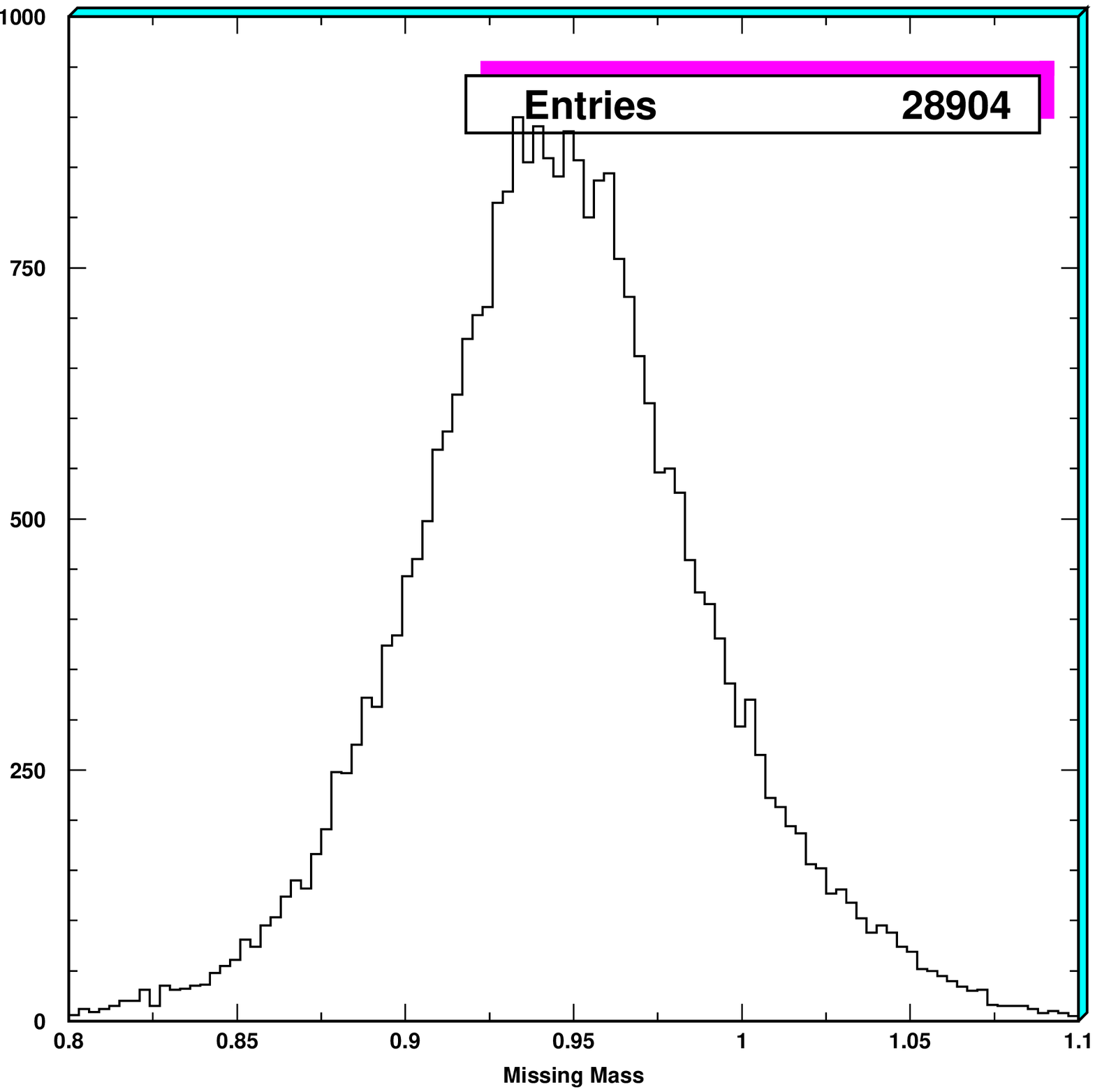,height=6.cm,angle=0,silent=} 
\hspace{0.5cm} \psfig{file=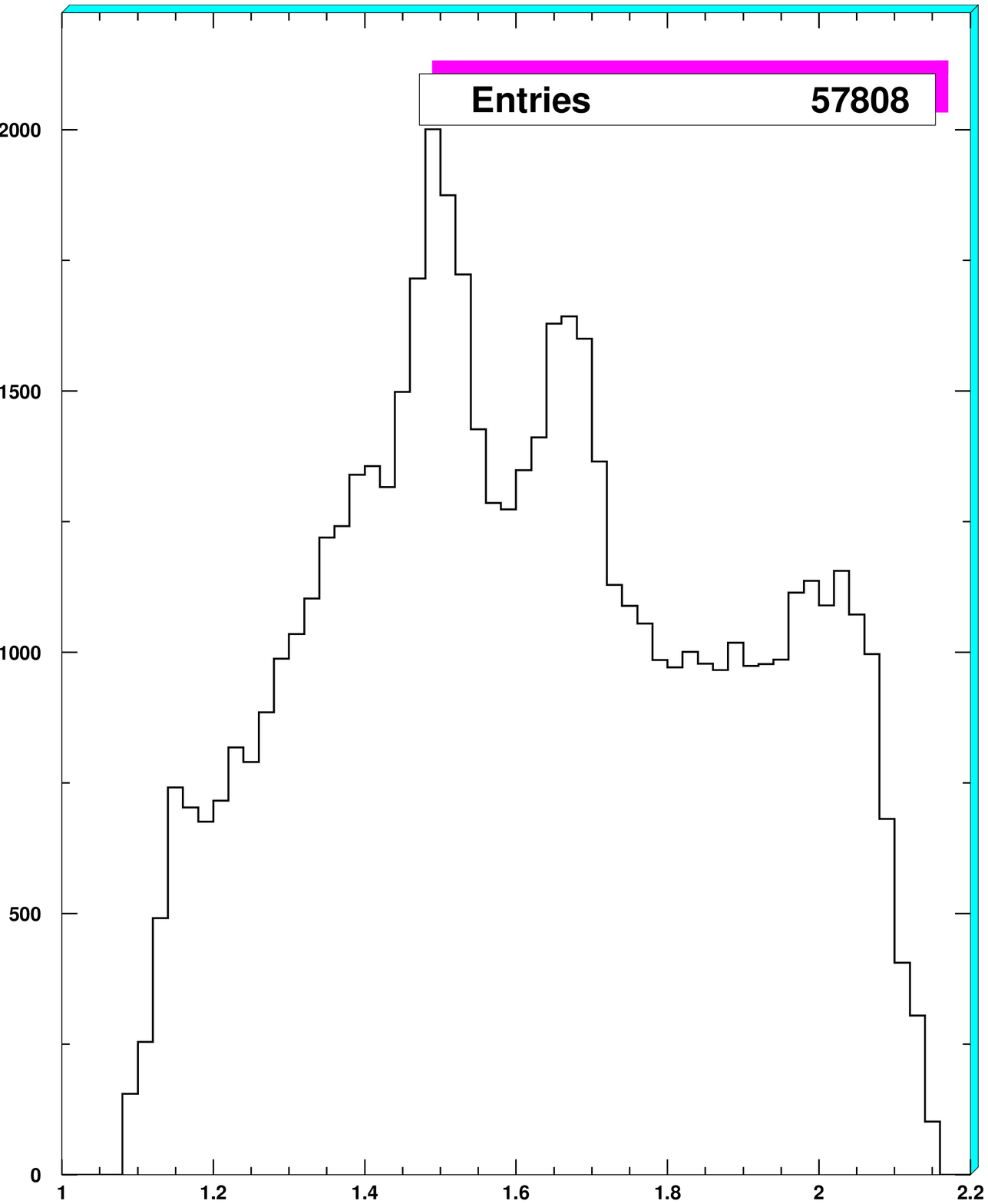,height=6.cm,
width=6.cm,angle=0,silent=}
}
\caption{\label{fig3} left: missing mass spectrum against $p\pi^-$ for
$J/\psi\to\bar np\pi^-$; right: $p\pi^-\&\bar n\pi^-$ invariant mass
spectrum for
$J/\psi\to\bar np\pi^-$. Preliminary BESII data}
\end{figure}

\begin{figure}[htbp]
\vspace{1.2cm}
\centerline{ \psfig{file=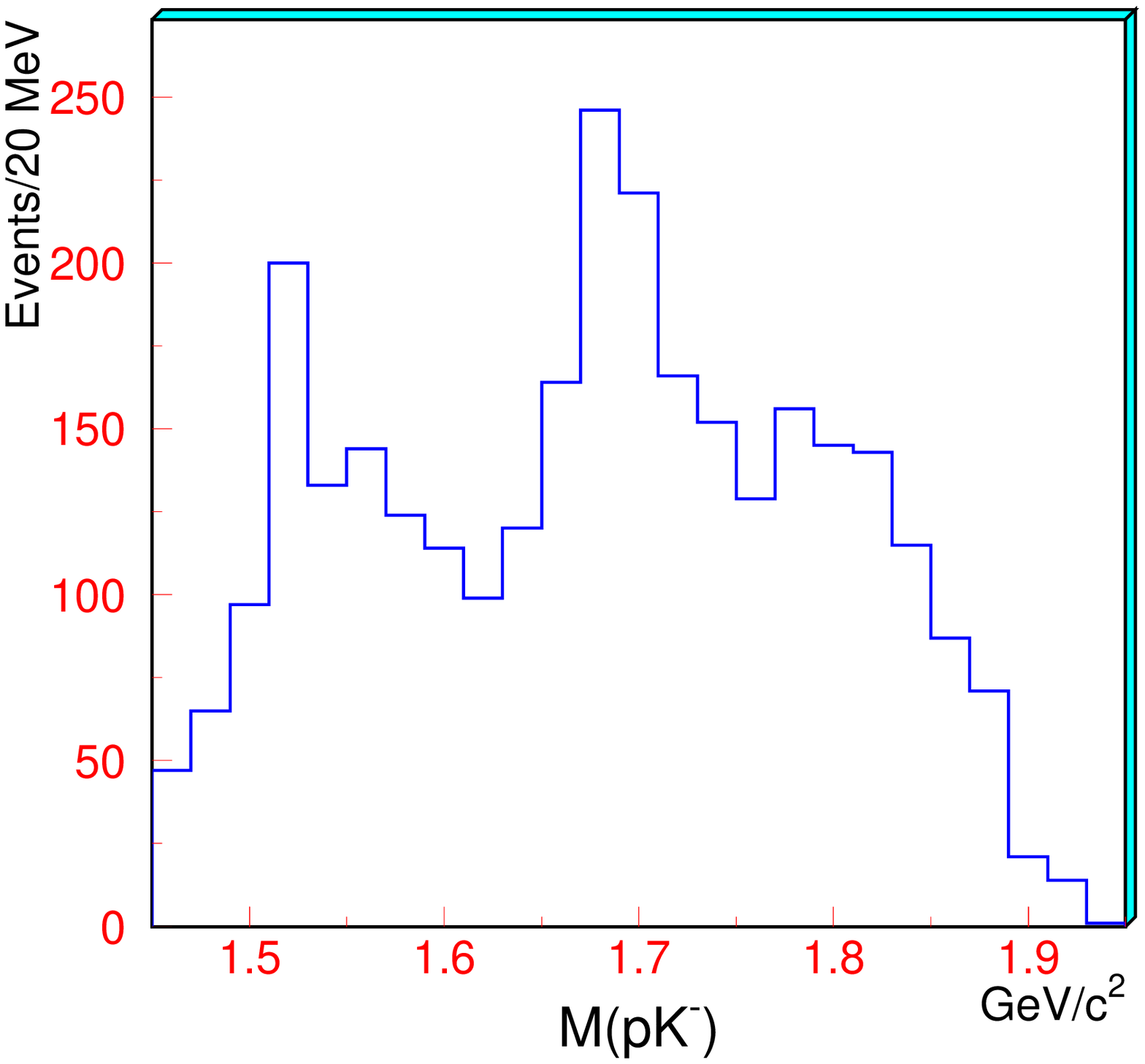,height=7.cm,angle=0,silent=}
\hspace{-0.7cm} \psfig{file=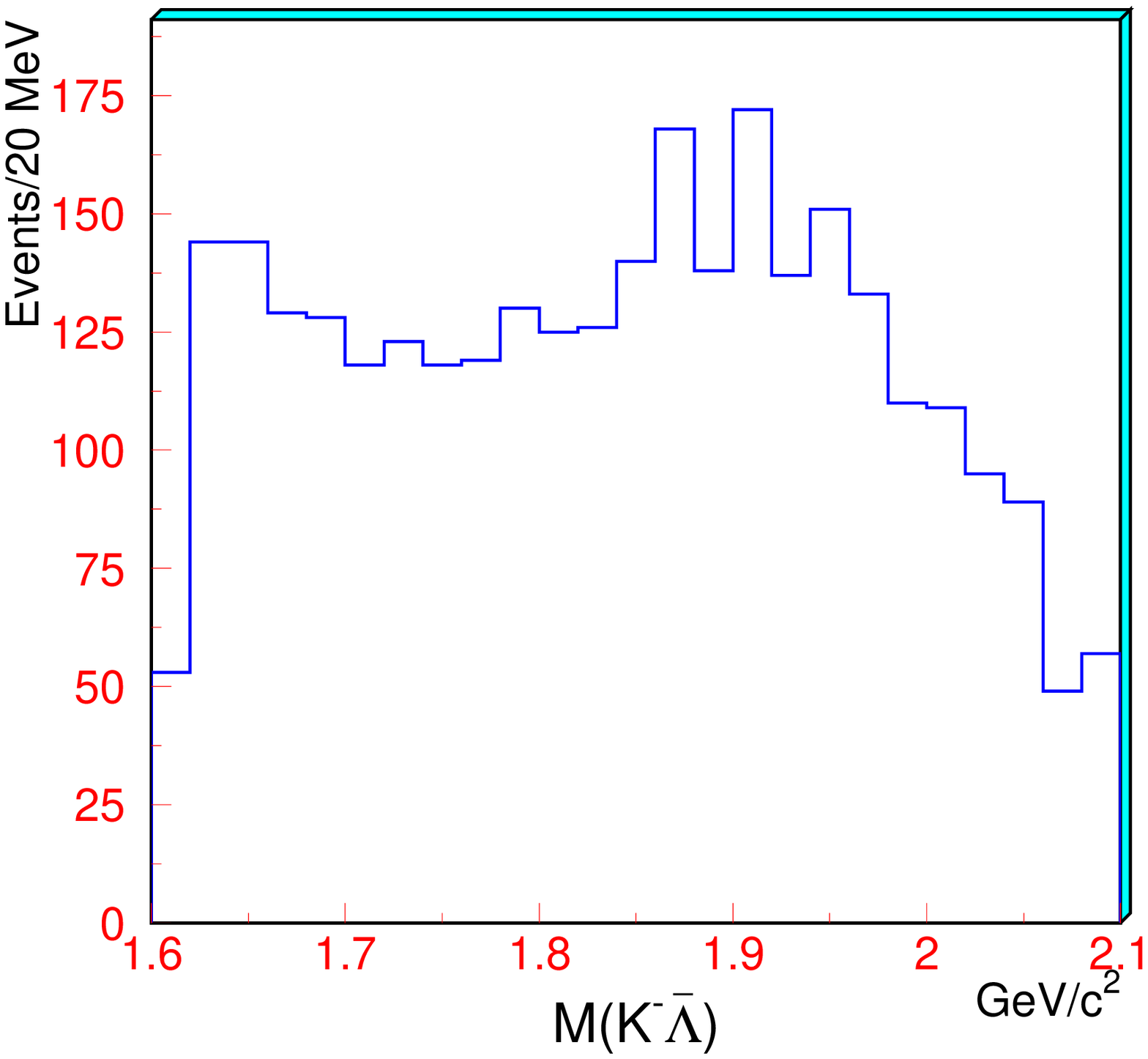,height=7.cm,angle=0,silent=}
}
\vspace{-2.cm}
\caption{\label{fig4} left: $pK$ invariant mass spectrum for
$J/\psi\to pK\Lambda$; right: $K\Lambda$ invariant mass
spectrum for
$J/\psi\to pK\Lambda$. Preliminary BESII data}
\end{figure}

With 58 million new $J/\psi$ events collected by BESII of improved
detecting efficiency, we have one order of magnitude more
reconstructed events for each channel. We show in
Figs.\ref{fig3} and \ref{fig4} preliminary results for $J/\psi\to p\bar
n\pi^-$  and $J/\psi\to pK^-\bar\Lambda + h.c.$ 
channels, respectively.

For $J/\psi\to p\bar n\pi^-$ channel, proton and $\pi^-$ are detected.
With some cuts of backgrounds, the missing mass spectrum shows a very
clean peak for the missing antineutron with negligible backgrounds;
The $N\pi$ invariant mass spectrum of
28,904 reconstructed events from half BESII
data looks similar to the $p\pi$ invariant mass spectrum
for $J/\psi\to p\bar p\pi^0$ as in Fig.~\ref{fig2}, but with much higher
statistics. Besides two very  clear peaks around 1500 and 1670 MeV,
the peak around 2020 MeV becomes clearer. This could be a ``missing"
$N^*$. For the decay $J/\psi\to\bar NN^*(2020)$,
the orbital angular momentum of $L=0$ is much preferred due to the
suppression of the centrifugal barrier factor for $L\leq 1$.
For $L=0$, the spin-parity of $N^*(2020)$ is limited to be $1/2+$ and   
$3/2+$. This may be the reason that the $N^*(2020) 3/2+$ shows up as  
a peak in $J/\psi$ decays while no peak shows up for $\pi N$ invariant 
mass spectra in $\pi N$ and $\gamma N$ production processes which allow
all $1/2\pm$, $3/2\pm$, $5/2\pm$ and $7/2\pm$ $N^*$ resonances
around 2.02 GeV to overlap and interfere with each other there. 

For $J/\psi\to pK^-\bar\Lambda$ and $\bar pK^+\Lambda$
channels, there are clear $\Lambda^*$ peaks at 1.52 GeV, 1.69 GeV and 1.8
GeV in $pK$ invariant mass spectrum, and $N^*$ peaks near $K\Lambda$
threshold and 1.9 GeV for $K\Lambda$ invariant mass spectrum. 
The SAPHIR experiment at ELSA\cite{SAPHIR,Bennhold} also observed a $N^*$
peak around 1.9 GeV for $K\Lambda$ invariant mass spectrum from photo-production.

We are also reconstructing $J/\psi\to\bar pp\omega$, $pK\Sigma$, $\bar pp\pi^+\pi^-$
and other channels. Partial wave analyses of various channels are in progress.

\section{$J/\psi$ radiative decays}

\begin{figure}[htbp]
\vspace{-1.6cm}
\hspace{-1.2cm}\includegraphics[width=15cm,height=6cm]{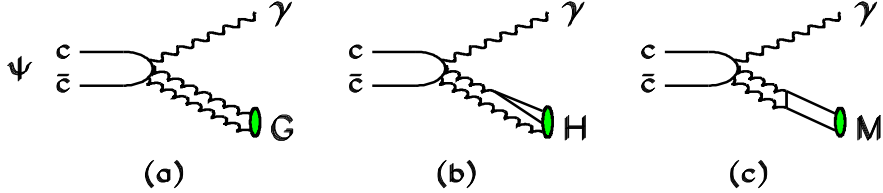}
\vspace{-1.8cm}
\caption{$\psi$ radiative decays to (a) glueball, (b)
hybrid, and (c) $q\bar q$ meson. }
\label{fig:5}
\end{figure}

There are three main physics objectives for $J/\psi$ radiative decays:

(1) Looking for glueballs and hybrids. As shown in Fig.~\ref{fig:5}, 
after emitting a photon, the $c\bar c$ pair is in a $C=+1$ state and
decays to hadrons dominantly through two gluon intermediate states.
Simply counting the power of $\alpha_s$ we know that glueballs should
have the largest production rate, hybrids the second,
then the ordinary $q\bar q$ mesons.

(2) Completing $q\bar q$ meson spectroscopy and studying their production
and decay rates, which is crucial for understanding their internal   
structure and confinement.

(3) Extracting $gg\leftrightarrow q\bar q$ coupling from perturbative
energy region of above 3 GeV to nonperturbative region of 0.3 GeV. 
This may show us some phenomenological pattern for the smooth transition
from perturbative QCD to strong nonperturbative QCD. 

Up to now, we have mainly worked on glueball searches. One thing worth
noting is that the $J/\psi$ radiative decay has a similar decay pattern as
$0^{-+}$, $0^{++}$ and $2^{++}$ charmoniums, {\sl i.e.}, $\eta_c$,   
$\chi_{c0}$ and $\chi_{c2}$, as it should be, since all of them decay
through two gluons. The $4\pi$, $\bar KK\pi\pi$, $\eta\pi\pi$ and
$\bar KK\pi$ seem to be the most favorable final states for the two gluon
transition at $1\sim 3$ GeV. The branching ratios for $J/\psi$ radiative decay to
these
four channels are listed in Table \ref{tab1}. The sum of them is about  
half of all radiative decays.  If glueballs exist, they should appear in
these four channels. Therefore BES Collaboration has performed
partial wave analyses (PWA) of these four
channels\cite{BES1,BES2,BES3,BES4} based on BESI data. The main results have been
summarized in Ref.~\cite{Zou3}. 
Mesons with large branching ratios in the $J/\psi$ radiative decays are
a very broad $\eta(2190)$ for $0^{-+}$, a broad $f_2(1950)$ for $2^{++}$,
$f_0(1500)$, $f_0(1710$-$1770)$ and $f_0(2100)$ for $0^{++}$. 

\begin{table}[htb]
\caption{Branching ratios for the four largest $J/\Psi$ radiative decay
channels (BR$\times 10^3$)}
\label{tab1}   
\renewcommand{\arraystretch}{1.2} %enlarge line spacing
\begin{center}
\begin{tabular}{cccc}
\hline
$\gamma 4\pi$ & $\gamma\bar KK\pi\pi$ & $\gamma\eta\pi\pi$ &
$\gamma\bar KK\pi$\\
\hline
$14.4\pm 1.8$ \cite{Kopke} & $9.5\pm 2.7$ \cite{BES2}
& $6.1\pm 1.0$ \cite{PDG} & $6.0\pm 2.1$ \cite{BES4} \\
\hline
\end{tabular}\\
\end{center}  
\end{table}

\begin{figure}[htbp]
%\vspace{1.2cm}
\centerline{ \psfig{file=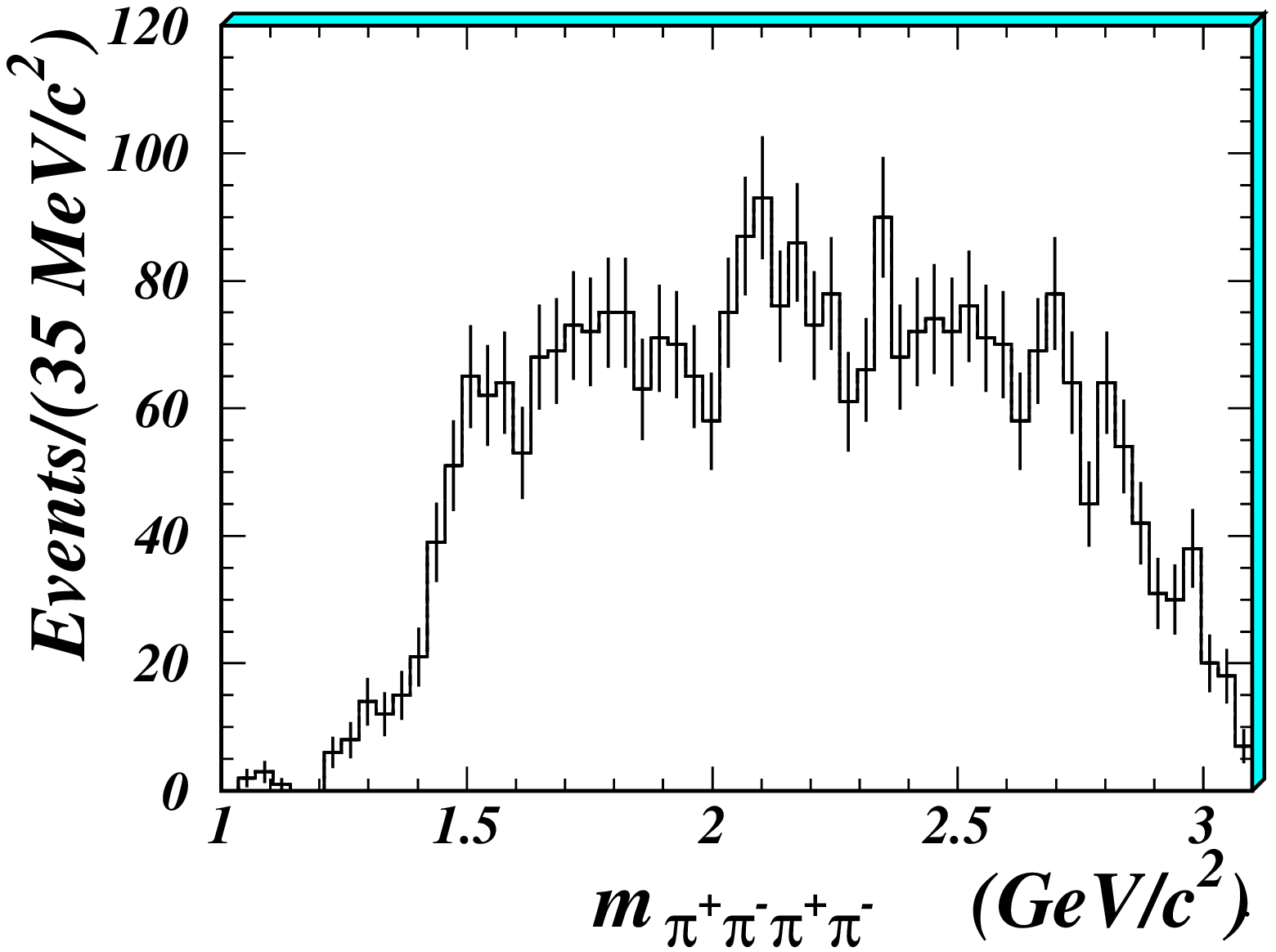,height=4.5cm,angle=0,silent=}
\hspace{-0.3cm} \psfig{file=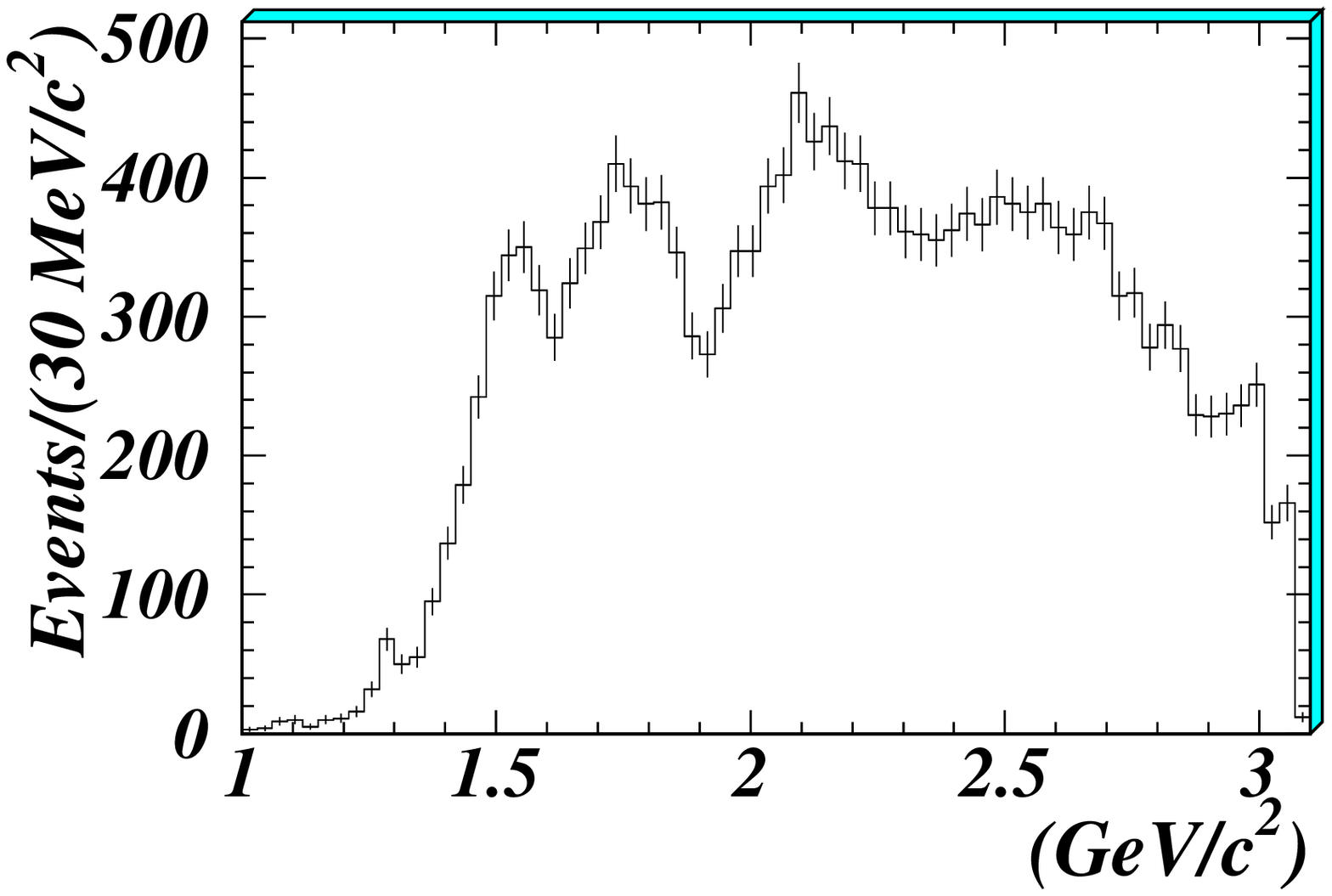,height=4.5cm,angle=0,silent=}
}
%\vspace{-2.cm}
\caption{\label{fig:g4pi} Comparison of BESI (left) and 
BESII data (right) for $J/\psi\to\gamma\pi^+\pi^-\pi^+\pi^-$.}
\end{figure}

With BESII data, all signals become clearer. For example, Fig.~\ref{fig:g4pi}
shows the comparison of BESI and BESII data for
$J/\psi\to\gamma\pi^+\pi^-\pi^+\pi^-$. 
For BESII data, we have performed partial wave analysis for the 
$\gamma\bar KK$  and $\gamma\pi^+\pi^-$ channels\cite{BES5}, where the main result
is that $f_J(1710)$ peak in these channels is definitely due to a $0^{++}$ particle.

\section{$J/\psi$ hadronic decays to mesons}

\begin{figure}[htbp]
\vspace{-1.6cm}
\hspace{-1.2cm}\includegraphics[width=15cm,height=6cm]{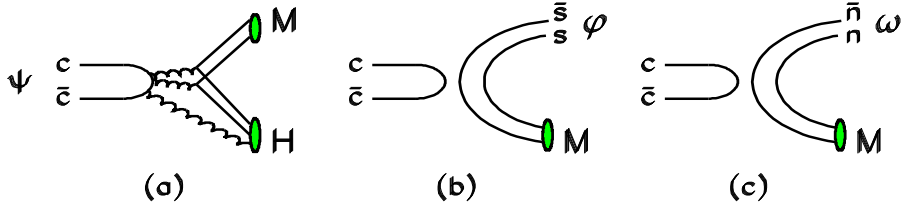}
\vspace{-1.8cm}
\caption{$\psi$ hadronic decays to (a) hybrids, (b)
$s\bar s$, and (c) $n\bar n\equiv {1\over\sqrt{2}}(u\bar u+d\bar d)$
mesons. }
\label{fig:2}
\end{figure}

There are mainly two physics objectives here:

(1) Looking for hybrids. Since $\psi$ decays to hadrons through three
gluons, final states involving a hybrid as shown in Fig.~\ref{fig:2}(a)
are expected to have larger production rate than ordinary $q\bar q$ mesons
as shown in Fig.~\ref{fig:2}(b,c).

(2) Extracting $u\bar u+d\bar d$ and $s\bar s$ components of
associated mesons, M, via $\Psi\to M+\omega/\phi$ as shown in
Fig.~\ref{fig:2}(b,c).

In order to look for isoscalar $1^{-+}$ hybrid $\hat\omega$ decaying to
$4\pi$, we have studied $J/\psi\to\omega\pi^+\pi^-\pi^+\pi^-$
process\cite{BES6}. A peak around 1.75 GeV in the $4\pi$ invariant mass
spectrum is visible. But due to low statistics, no PWA is performed.
No other structure is observed.

\begin{figure}[htbp]
\vspace{5.2cm}
\includegraphics{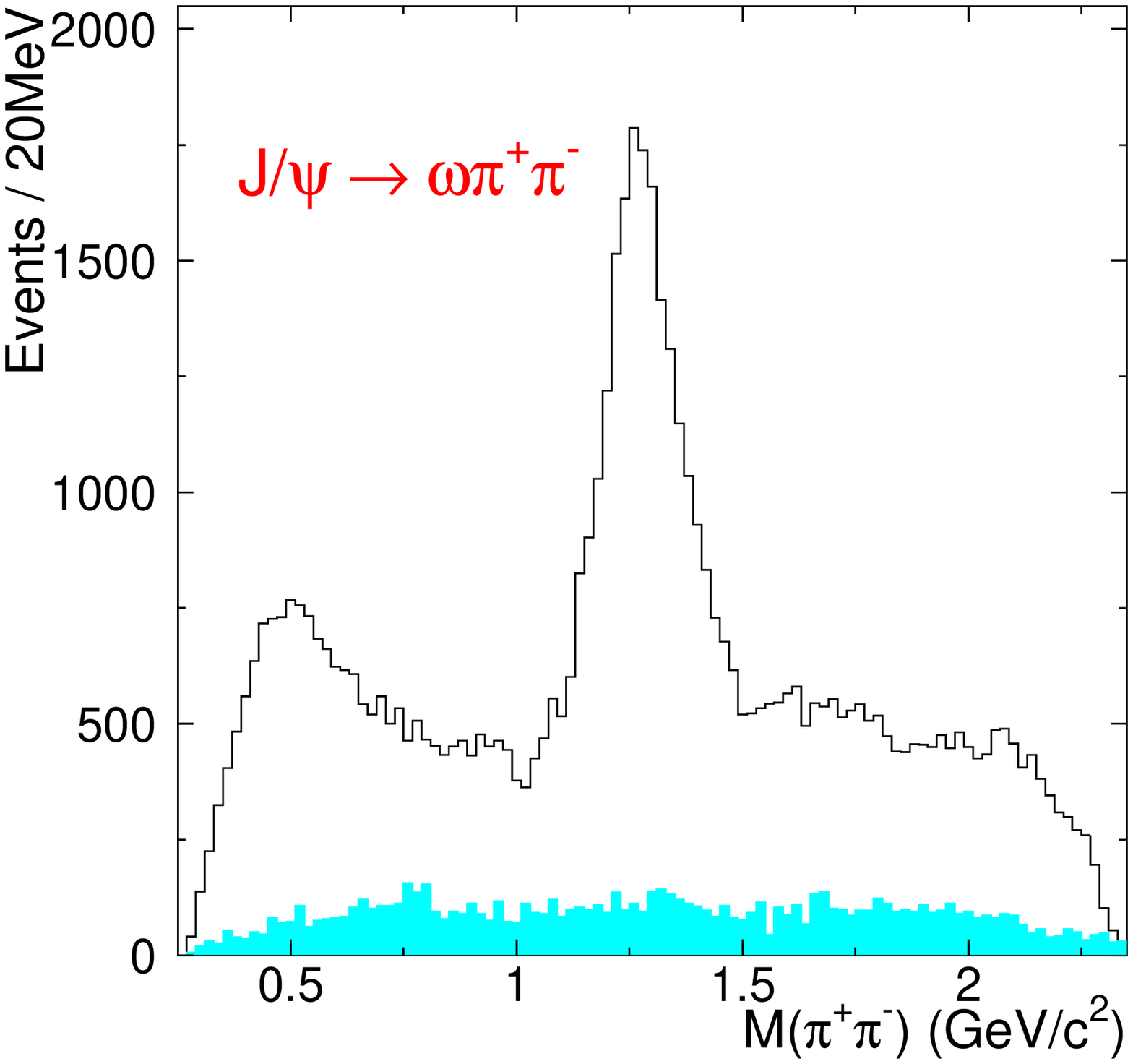}
\includegraphics{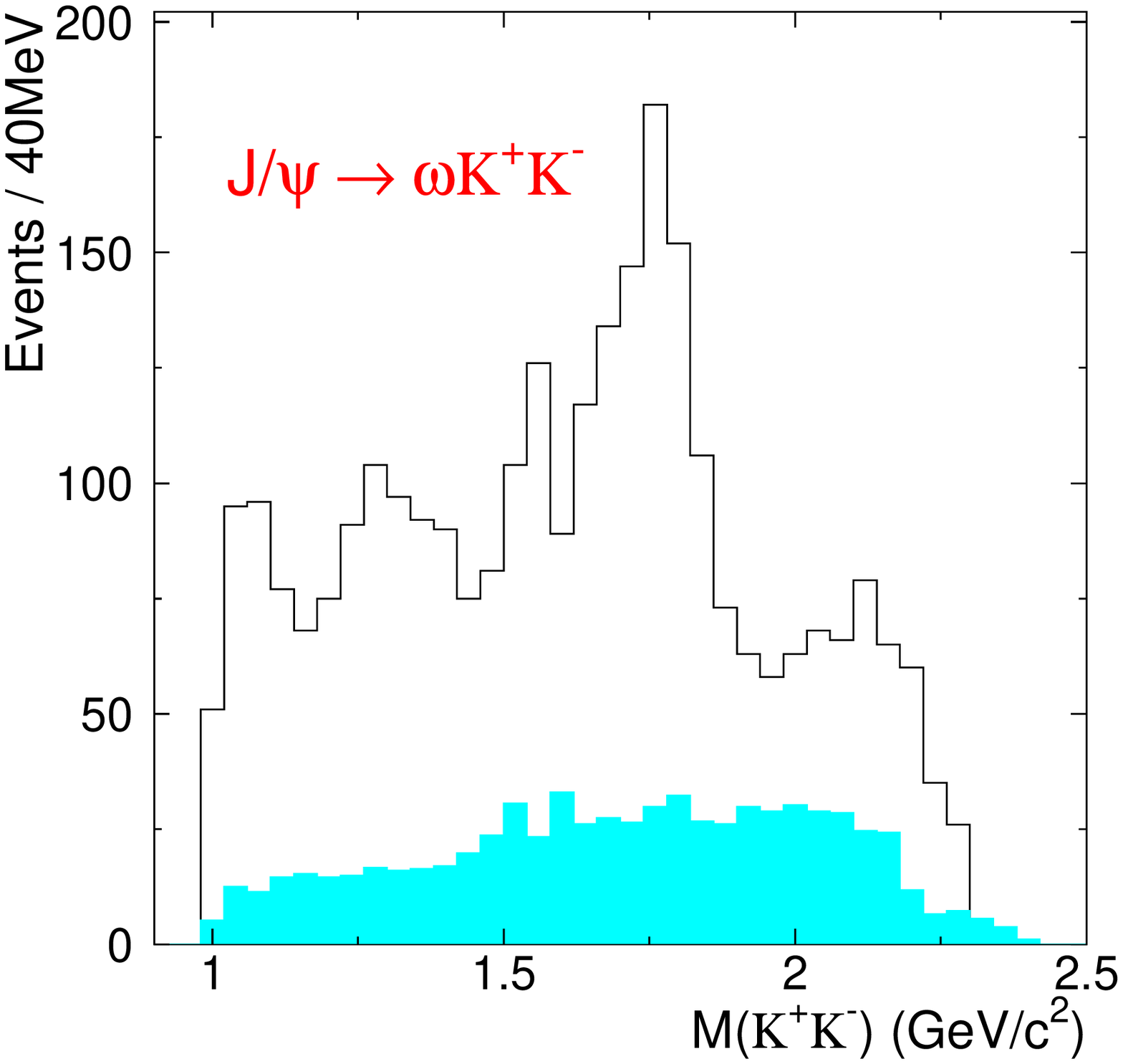}
\caption{\label{fig:wpp} left: $\pi\pi$ invariant mass spectrum for
$J/\psi\to\omega\pi^+\pi^-$; right: $K^+K^-$ invariant mass
spectrum for
$J/\psi\to\omega K^+K^-$. Preliminary BESII data}
\end{figure}

\begin{figure}[htbp]
\vspace{0.2cm}
\centerline{ \psfig{file=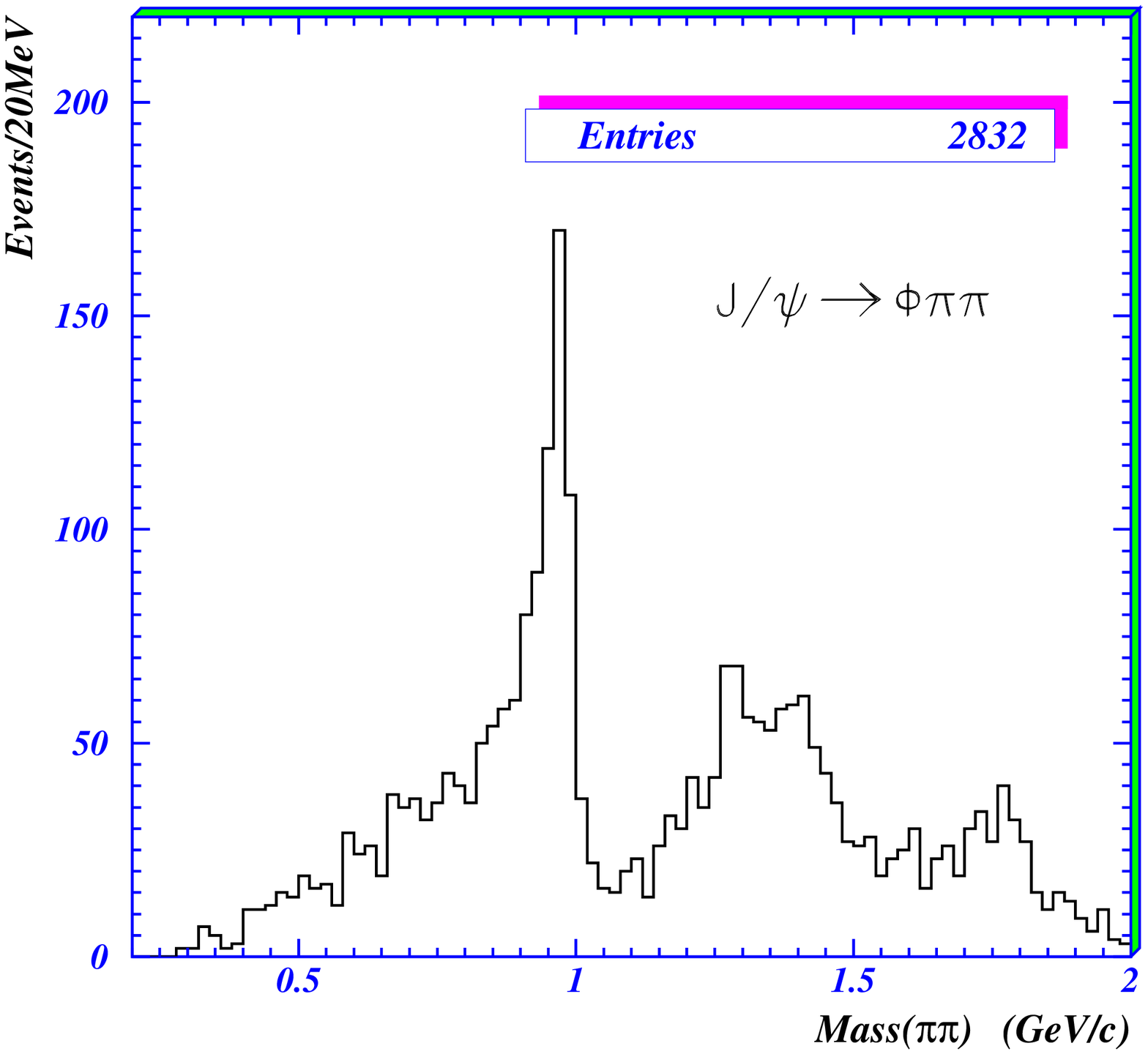,height=5.5cm,angle=0,silent=}
\hspace{0.4cm} \psfig{file=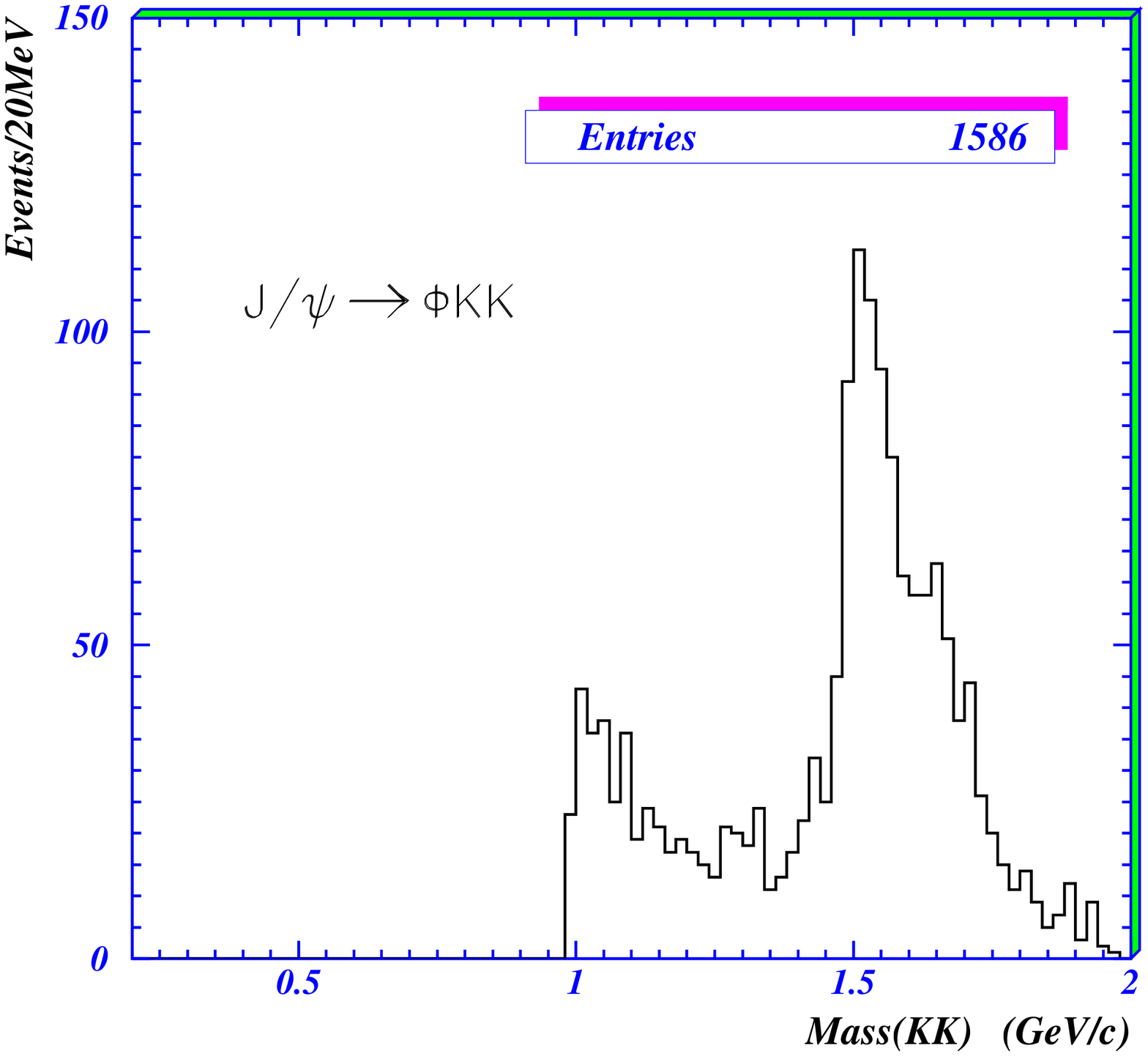,height=5.5cm,angle=0,silent=}
}
%\vspace{-2.cm}
\caption{\label{fig:fpp} left: $\pi\pi$ invariant mass spectrum for
$J/\psi\to\phi\pi^+\pi^-$; right: $K^+K^-$ invariant mass
spectrum for $J/\psi\to\phi K^+K^-$. Preliminary half BESII data}
\end{figure}

To investigate the $u\bar u+d\bar d$ and $s\bar s$ components of mesons,
we have studied $J/\psi\to\omega\pi^+\pi^-$, $\omega K^+K^-$,
$\phi\pi^+\pi^-$ and $\phi K^+K^-$ channels. The invariant mass spectra
for these channels are shown in Fig.~\ref{fig:wpp} and Fig.~\ref{fig:fpp}, which are
similar to the previous ones by MARKIII and DM2
Collaborations, but with much higher statistics.

For $J/\psi\to\omega\pi^+\pi^-$, there are two clear peaks at 500 MeV and
1275 MeV in the $2\pi$ mass spectrum corresponding to the $\sigma$  
and the $f_2(1275)$, respectively\cite{Shen}.
For $J/\psi\to\omega K^+K^-$, there
is a threshold enhancement due to the $f_0(980)$ and a clear peak at 1710
MeV probably due to the $f_0(1710)$.

Preliminary results\cite{Shen} of partial wave analyses indicate that
(1) in the $\pi\pi$ mass spectrum of the $J/\psi\to\phi \pi^+\pi^-$
process all three peaks at 980 MeV, 1330 MeV and 1770 MeV are dominantly
$0^{++}$; (2) in the $K\bar K$ mass spectrum of the $J/\psi\to\phi K^+K^-$
the peak at 1525 MeV is due to $f'_2(1525)$ while the $K\bar K$ threshold
enhancement and the shoulder around 1700 MeV are due to $f_0(980)$ and
$f_0(1710)$, respectively. 

In summary, the $\sigma$ and $f_2(1275)$
appear clearly only in the $J/\Psi\to\omega+X$ process, the $f'_2(1525)$
appears clearly only in the $J/\Psi\to\phi+X$ process, and $f_0(980)$
and $f_0(1710$-$1770)$ appear clearly in both processes.

\section{Future prospects}

We are now working on the 58 million $J/\psi$ events collected with BESII detector 
in the years from 1999 to 2001. Physics results on various channels of $N^*$ and
meson production are expected to be published in near future.

We have been taking $\psi'(3686)$ data since last year and hope to reach more than
20 million $\psi'$ events in next year. The data of $\psi'$ decays will extend 
our study on $N^*$ and meson resonances to a broader energy range.

A major upgrade of the collider to BEPCII is planned to be finished in
about 4 years. A further two order of magnitude more
statistics is expected to be achieved. Such statistics will
enable us to perform partial wave analyses of plenty important channels
for both meson spectroscopy and baryon spectroscopy from the $J/\psi$ and
$\psi'$ decays.
We expect BEPCII to play a very important role in many aspects
of light hadron spectroscopy, such as hunting for the glueballs and
hybrids, extracting $u\bar u+d\bar d$ and $s\bar s$ components of mesons,
and studying excited nucleons and hyperons, {i.e.}, $N^*$, $\Lambda^*$,
$\Sigma^*$ and $\Xi^*$ resonances.
 
\section*{Acknowledgments}
We would like to thank the workshop organizers for
their kind invitation and financial support for our participation of
this very interesting and successful conference. This work is partly
supported by the CAS Knowledge Innovation Project (KJCX2-N11) and 
National Science
Foundation of China, and is written up during author's visit at ICTP, 
Trieste, Italy.

\end{document}